\documentclass[prl,twocolumn,superscriptaddress,preprintnumbers,amsmath,amssymb]{revtex4-1}
\usepackage{graphicx}
\usepackage{color}
\usepackage{bm}
\usepackage{ulem} 

\begin{document}

\title{Dephasing and dissipation in qubit thermodynamics}

\author{J. P. Pekola}
\affiliation{Low Temperature Laboratory, Department of Applied Physics, Aalto University School of Science, P.O. Box 13500, 00076 Aalto, Finland}
\author{Y. Masuyama}
\affiliation{Research Center for Advanced Science and Technology (RCAST), The University of Tokyo, Meguro-ku, Tokyo 153-8904, Japan}
\author{Y. Nakamura}
\affiliation{Research Center for Advanced Science and Technology (RCAST), The University of Tokyo, Meguro-ku, Tokyo 153-8904, Japan}
\affiliation{RIKEN Center for Emergent Matter Science (CEMS), Wako, Saitama 351-0198, Japan}
\author{J. Bergli}
\affiliation{Physics Department, University of Oslo, P.O.Box 1048 Blindern, 0316 Oslo, Norway}
\author{Y. M. Galperin}
\affiliation{Physics Department, University of Oslo, P.O.Box 1048 Blindern, 0316 Oslo, Norway}
\affiliation{A. F. Ioffe Physico-Technical Institute RAS, 194021 St.
Petersburg, Russian Federation}

\begin{abstract}
We analyze the stochastic evolution and dephasing of a qubit within the quantum jump (QJ) approach. It allows one to treat individual realizations of inelastic processes, and in this way it provides solutions, for instance, to problems in quantum thermodynamics and distributions in statistical mechanics. As a solvable example, we study a qubit in the weak dissipation limit, and demonstrate that dephasing and relaxation render the Jarzynski and Crooks fluctuation relations (FRs) of non-equilibrium thermodynamics intact. On the contrary, the standard two-measurement protocol, taking into account only the fluctuations of the internal energy $U$, leads to deviations in FRs under the same conditions. We relate the average $\langle e^{-\beta U} \rangle $  (where $\beta$ is the inverse temperature) with the qubit's relaxation and dephasing rates, and discuss this relationship for different mechanisms of decoherence.
\end{abstract}

\date{\today}

\maketitle
Decoherence of quantum systems is a topic of considerable fundamental interest, but it also presents an important technological challenge in the pursuit of quantum computing based on two-level systems (TLSs), qubits, see, e.~g.,~\cite{Chirolli2008} and references therein. 
Quite generally decoherence is considered on the level of averages over many experiments: in this situation results predicting expectation values can be obtained theoretically, for instance, using master equations for the density matrix of the system.  It is, however, of considerable interest to obtain direct access to decoherence processes  from the point of view of non-equilibrium statistical mechanics and thermodynamics for the system and its environment. This problem can be tackled conveniently by various approaches of stochastic quantum mechanics, e.g., via the analysis of quantum jumps (QJ) and non-Hermitian Schr\"odinger equation~\cite{Dalibard92}. It is the purpose of this article to include  dephasing in such analysis, and to discuss various models of decoherence from the thermodynamic point of view.
\begin{figure}[t]
    \begin{center}
    \includegraphics[width=.8\columnwidth]{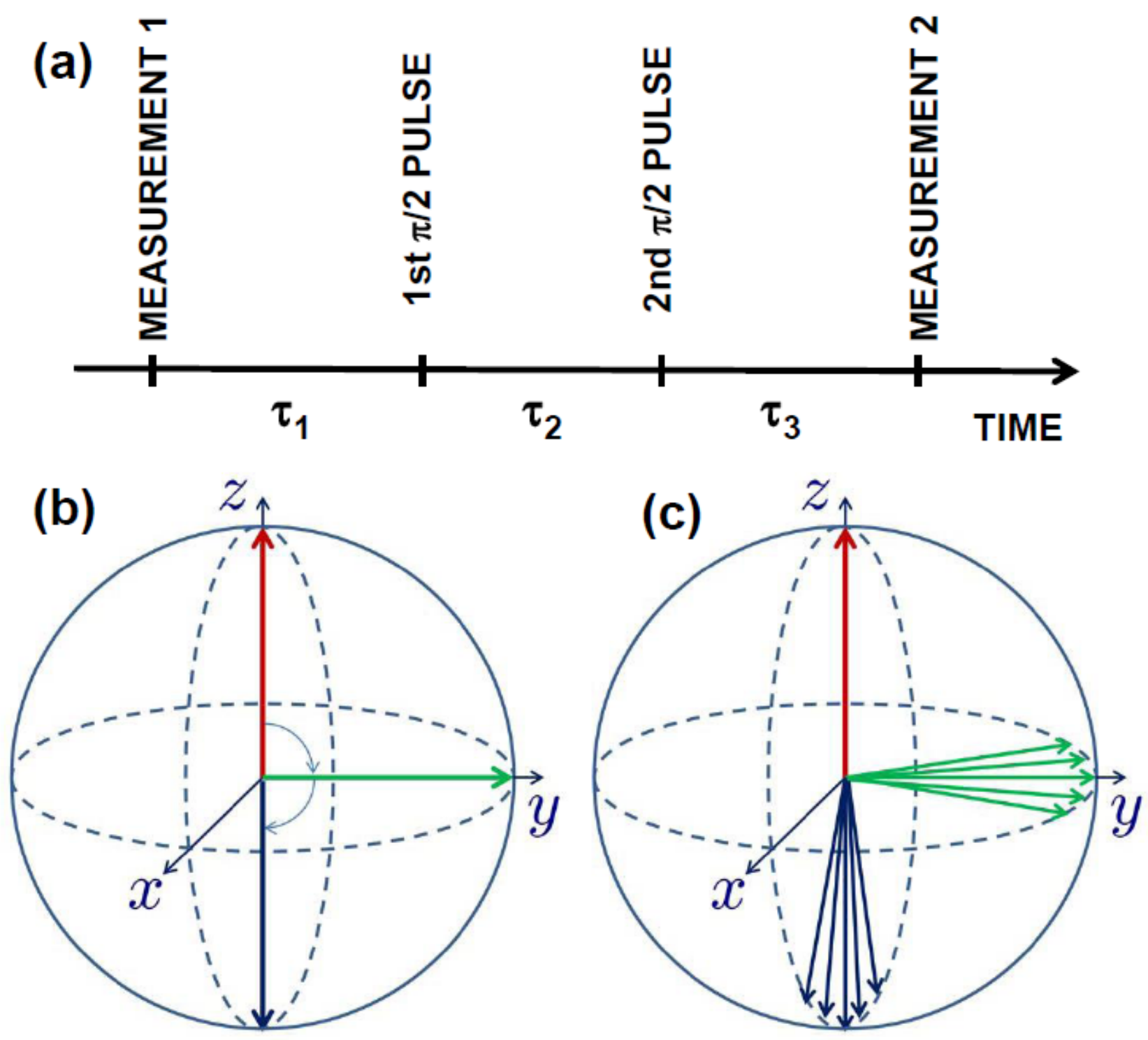}
    \end{center}
    \caption{\label{fig1} (a) The pulses applied to the qubit in TMP protocol. The qubit state is measured in the beginning and at the end. In between, two $\pi/2$ pulses around $x$-axis are applied. The three time intervals between the pulses are denoted $\tau_1,\tau_2$ and $\tau_3$. The corresponding evolution of the qubit is shown on the Bloch sphere for two consequent $\pi/2$-pulses ($\tau_2=0$) in panel (b), and in the presence of dephasing during the time interval $\tau_2$ in panel (c).}
\end{figure}   

Work $W$ done on a system by a source splits into the ``useful" work, which is the change $U$ in the internal energy of the system itself, and into heat dissipated into the environment, $Q$, such that 
\begin{equation} \label{w1}
W=U+Q. 
\end{equation}
In order to relate such a process, for instance, to the thermodynamic fluctuation relations~\cite{Bochkov81,Jarzynski97,Crooks99}, 
see~\cite{Esposito09,Campisi11} for a review, 
the full work ($W$) needs generally to be considered. There are various ways to measure work-related quantities in a quantum set-up. The original proposal was put forward as the so-called two-measurement protocol (TMP), where the state of the system is measured first before the work is applied, and second after the application of this work \cite{Kurchan00,Talkner07}. This yields naturally the difference in the internal energy in form $U=E_f -E_i$, where $E_i,E_f$ refer to the energies of the states of the system observed in the initial and final measurements, respectively. For a closed system, not interacting with the environment during the driving period ($Q=0$), this yields then the whole work according to Eq.~\eqref{w1}. The true interesting case is, however, that where the system is open ($Q\neq 0$ in general), which cannot be captured by the simple TMP. Because TMP, being an obvious first choice, has nevertheless become a common measure of work in actual experiments on quantum 
systems~\cite{An15}, see also~\cite{Batalh14}, 
it is interesting to see quantitatively how good or bad an estimate it yields for the full work in realistic open set-ups.
To perform such an analysis is one of the goals of this article, and in quantitative terms we present, based on the QJ method \cite{Hekking13}, results on fluctuation relations on $U$ for a generic two-level system with definite environmental relaxation/excitation rates and dephasing at finite temperatures. In the case of weak dissipation, we obtain leading order corrections analytically.

To obtain the governing equation, we write the evolution of the wave-function of a two-level system in the scenario which accounts for driving, relaxation and dephasing. In the absence of jumps the wave-function obeys a Schr\"odinger like equation with a non-Hermitian Hamiltonian $H$ such that \cite{Dalibard92}
\begin{equation} \label{g0}
|\psi (t+ \Delta t)\rangle = (1-p)^{-1/2} \left(1 - \frac{i}{\hbar} H \Delta t \right)|\psi (t)\rangle,
\end{equation}  
where $|\psi(t)\rangle=a(t)|g\rangle+b(t)|e\rangle$, $p= (\Gamma_\uparrow |a|^2 + \Gamma_\downarrow |b|^2)\Delta t$ for a small time interval of observation,  $\Delta t$. Here, $\Gamma_\downarrow,\Gamma_\uparrow$ are the relaxation and excitation rates, respectively, due to the coupling to the bath, and they obey the detailed balance condition $\Gamma_\uparrow = e^{-\beta \hbar\omega_0}\Gamma_\downarrow$ at inverse temperature $\beta$ for level spacing $\hbar\omega_0$ of the qubit. These rates determine the probabilities of interrupting the evolution of Eq.~\eqref{g0} by excitation, $p_\uparrow = \Gamma_\uparrow |a|^2 \Delta t$, or by relaxation, $p_\downarrow = \Gamma_\downarrow |b|^2 \Delta t$, during the time interval $\Delta t$.
$|g \rangle$ and $|e \rangle$ denote the qubit ground and excited state, respectively.
In the interaction picture  we take the unperturbed Hamiltonian $H_0$ of a TLS  with the level spacing  equal to the drive frequency $\omega_0$. Thus, we may write the total non-Hermitian Hamiltonian (excluding that of the bath itself) of the quantum trajectory approach as
\begin{equation} \label{g1}
H= H_0 + \delta H + V + N.
\end{equation}  
Here $H_0 = (\hbar \omega_0/2) (|e\rangle\langle e| - |g\rangle\langle g|)$. The slow fluctuation of the level spacing responsible for dephasing reads $\delta H = (\hbar \delta \omega /2) (|e\rangle\langle e| - |g\rangle\langle g|)$.
We assume that $\delta \omega (t)\ll \omega_0$ is a classical stochastic process. The driving term, with the drive signal $\lambda(t)$ is chosen to be of the form $V = \lambda(t) (|g\rangle\langle e| + |e\rangle\langle g|)$,
and the non-Hermitian part of the Hamiltonian reads
$N = -(i\hbar/2) (\Gamma_\downarrow |e\rangle\langle e| + \Gamma_\uparrow |g\rangle\langle g|)$.
We write the Schr\"odinger  equation  for the wave-function in the interaction picture,
$|\psi_I (t)\rangle = e^{i H_0  t/\hbar }|\psi (t)\rangle$.   Then expanding up to linear order in $\Delta t$, we find the evolution corresponding to Eq.~\eqref{g0}:
\begin{eqnarray} \label{g7}
&& |\psi_I (t+ \Delta t)\rangle =(1-p)^{-1/2}\times\nonumber\\
&& \quad 
\left[1 - \frac{i}{\hbar} (V_I(t) + \delta H_I (t) + N_I) \Delta t \right]|\psi_I (t)\rangle,
\end{eqnarray}  
where operator $\mathcal O$ in the interaction picture reads $\mathcal O_I (t) = e^{i H_0  t/\hbar }\mathcal O e^{- i H_0  t/\hbar }$. 
For the two state system, the evolution of $a$ and $b$ of $|\psi_I (t)\rangle = a |g\rangle + b|e\rangle$ obeys then
\begin{eqnarray} \label{g8}
&& \dot a = -\frac{i}{\hbar} \lambda(t) e^{-i\omega_0 t} b + i\frac{\delta \omega}{2} a + \frac{\Delta \Gamma}{2}a|b|^2, \nonumber \\ && \dot b = -\frac{i}{\hbar} \lambda(t) e^{i\omega_0 t} a - i\frac{\delta \omega}{2} b - \frac{\Delta \Gamma}{2}|a|^2 b.
\end{eqnarray}
Here we have defined $\Delta\Gamma \equiv \Gamma_\downarrow - \Gamma_\uparrow$.

In what follows, we analyze an exemplary protocol in the weak dissipation regime, and demonstrate that, as expected, the Jarzynski equality (JE)
\begin{equation} \label{je}
\langle e^{-\beta W}\rangle =1
\end{equation}
holds, whereas $\langle e^{-\beta U}\rangle$ differs from unity. Before presenting the formal derivation of the weak dissipation results, we justify such a deviation by a simple argument in the limit where the time interval between the two measurements is very long. In general, we can write the TMP outcome as
\begin{eqnarray} \label{hh1}
\langle e^{-\beta U}\rangle = && \, p_g^i\, p^f_{g|g}\, e^0 + p_g^i\, p^f_{e|g}\, e^{-\beta\hbar\omega_0} 
\nonumber\\ &&+ p_e^i\, p^f_{g|e}\, e^{\beta\hbar\omega_0} + p_e^i\, p^f_{e|e}\, e^0.
\end{eqnarray}
Here $p^i_a$ is the probability of detecting state $a=g,e$ at the first measurement and $p^f_{b|a}$ is the conditional probability of measuring state $b$ at the end (2nd measurement, $f$), if the system was initially in state $a$. Taking the time interval between the two measurements to be very long means that all the populations in Eq.~\eqref{hh1} are thermally distributed, $p_g^i = 1- p_e^i = p^f_{g|g,e}=1-p^f_{e|g,e}= (1+e^{-\beta\hbar\omega_0})^{-1}$, and Eq. \eqref{hh1} yields
\begin{equation} \label{hh2}
\langle e^{-\beta U}\rangle =  1+\tanh^{2} (\beta\hbar\omega_0/2).
\end{equation}
In the high-temperature limit ($\beta \hbar\omega_0 \ll 1$), this expression yields $\langle e^{-\beta U}\rangle =1$, but at low temperatures ($\beta \hbar\omega_0 \gg 1$), we obtain  $\langle e^{-\beta U}\rangle = 2$ in stark contradiction with JE. This result holds for any driving protocol between the two measurements.

Now we move to the weak dissipation treatment 
taking into account trajectories which include at most one relaxation or excitation event. This way we obtain deviations from the 
fluctuation relations linear in the transition rates. We may write for the work exponent
\begin{equation} \label{h2}
\langle e^{-\beta W}\rangle = P_0\langle e^{-\beta W}\rangle_0 + P_1 \langle e^{-\beta W}\rangle_1,
\end{equation}   
where $P_0,P_1$ are the probabilities of zero and one-photon processes, and $\langle e^{-\beta W}\rangle_0,\langle e^{-\beta W}\rangle_1$ are the averages for the corresponding processes. We will explicitly check the validity of JE up to this order. On the other hand, we have for the quantity measured in the TMP protocol,
\begin{equation} \label{h3}
\langle e^{-\beta U}\rangle = P_0\langle e^{-\beta U}\rangle_0 + P_1 \langle e^{-\beta U}\rangle_1
\end{equation}  
with the corresponding notations. Now for zero-photon processes, $U=W$, thus $\langle e^{-\beta U}\rangle_0 = \langle e^{-\beta W}\rangle_0$.

We choose the following protocol, see Fig.~\ref{fig1}. The qubit is measured in the beginning and at the end of the process (TMP), and it is driven in between by two consecutive pulses, each of them producing ideally a $\pi/2$ rotation around the same, say $x$-axis. We assume that the pulses are short enough in time such that no relaxation or dephasing takes place during them. This is possible to achieve by adjusting the drive amplitude of the rotation pulses. The timing outside the pulses is such that the first $\pi/2$ pulse is applied at time $\tau_1$ after the initial measurement of the qubit state. The time interval between the two rotation pulses is denoted $\tau_2$, and during this time dephasing can take place. Finally, $\tau_3$ is the time interval between the second rotation pulse and the second measurement pulse. 
This sequence has the property that without relaxation and dephasing it performs a standard $\pi$ rotation swapping the system between the states $g$ and $e$. Yet, the qubit may jump at any moment during the protocol, if $\Gamma$'s are finite, and in between the two $\pi/2$ pulses the qubit is in a superposition state and it is susceptible to dephasing as well. Typical trajectories on the Bloch sphere are illustrated in Fig.~\ref{fig1}~(b) for $\tau_2=0$ ($\pi$ pulse) and in (c) for $\tau_2 > 0$ (dephasing) starting in the state $g$ at the north pole.

Derivation of the averaged work exponents \eqref{h2} and \eqref{h3} is made along the following lines. We follow the amplitudes along the operation of the qubit according to the protocol shown in Fig.~\ref{fig1}. We take into account those trajectories that involve either no or one quantum jump. Knowing the amplitudes we evaluate the probabilities of different trajectories and then calculate the thermal averages up to the first order in $\Gamma$'s. The detailed calculation, outlined in the Supplemental Material, Sec.~1,  results in the relationship $\langle e^{-\beta W}\rangle_{\delta\omega} =1$
where the subscript $\delta\omega$ emphasizes that we average over all realizations with a fixed level spacing $\omega_0+\delta\omega$. This result implies that the Jarzynski equality is valid for any distribution of $\delta \omega$, $\langle e^{-\beta W}\rangle =1$. 
On the contrary, for $U$ we obtain
\begin{equation} \label{ju1}
\langle e^{-\beta U}\rangle_{\delta\omega} -1 = [\tau_3 - \tau_1 \cos(\delta\varphi_2)]\,\Gamma_\Sigma \tanh^2 (\beta\hbar\omega_0/2) 
\end{equation} 
where $\Gamma_\Sigma \equiv \Gamma_\downarrow + \Gamma_\uparrow$ and  $\delta \varphi_2 \equiv \int_{\tau_1}^{\tau_1+\tau_2}\! \!  \delta \omega (t)\, dt$.
After averaging over realizations of the random process   $\delta \omega (t)$ we get
\begin{equation} \label{yu1}
\langle e^{-\beta U}\rangle -1 = [\tau_3 - \tau_1  \langle \cos(\delta\varphi_2) \rangle]\,\Gamma_\Sigma \tanh^2 (\beta\hbar\omega_0/2) .
\end{equation} 

Equation~\eqref{yu1} is the central result of the present work -- it relates the difference $\Delta \equiv \langle e^{-\beta U}\rangle -1$ with decoherence of the qubit characterized by the average  $ \langle \cos(\delta\varphi_2) \rangle$. 
If the qubit is fully characterized and the above average is known, then Eq.~\eqref{yu1} can be used for checking the JE. Indeed,
since $\langle e^{-\beta W}\rangle =1$ we can correct for usage of $U$ instead of $W$. If $\Gamma_\Sigma (\tau_1+\tau_2+\tau_3) \ll 1$, then the correction is negligible.

On the other hand, if the quantity $\Delta (\{\tau_i\})$ is determined from the experiment (say, by direct registration of emitted or absorbed photons), then an information regarding dephasing can we extracted as
\begin{equation}\label{yu1a}
\langle \cos (\delta \varphi_2)\rangle=\frac{\tau_3}{\tau_1}\left[ 1 - \left(\frac{\partial \ln \Delta}{\partial \ln \tau_3}\right)^{-1}\right]\, .
\end{equation}
The quantity $\langle \cos (\delta \varphi_2)\rangle$  contains the same information as what can be obtained in a Ramsey 
measurement~\cite{Ithier05}. This quantity appears due to the specially selected protocol shown in Fig.~\ref{fig1}. Other protocols would contain different averages, and in this way may provide an additional information about the underlying decoherence mechanism. For instance, when a $\pi/4$-rotation is first performed around $y$-axis, and then a $\pi/2$-rotation around $x$-axis, one obtains corrections $\propto\langle \sin (\delta \varphi_2)\rangle$.

In the absence of dephasing $\langle \cos (\delta \varphi_2)\rangle=1$ and
\begin{equation} \label{yu1b}
  \langle e^{-\beta U}\rangle -1 = (\tau_3 - \tau_1) \Gamma_\Sigma \tanh^2 (\beta\hbar\omega_0/2) .
\end{equation}
The result depends on $(\tau_3-\tau_1)$ through the difference of the probabilities of photon-assisted processes during the time slots $\tau_3$ and $\tau_1$, respectively.
\begin{figure}[b]
\centering
\includegraphics[width=.7\columnwidth]{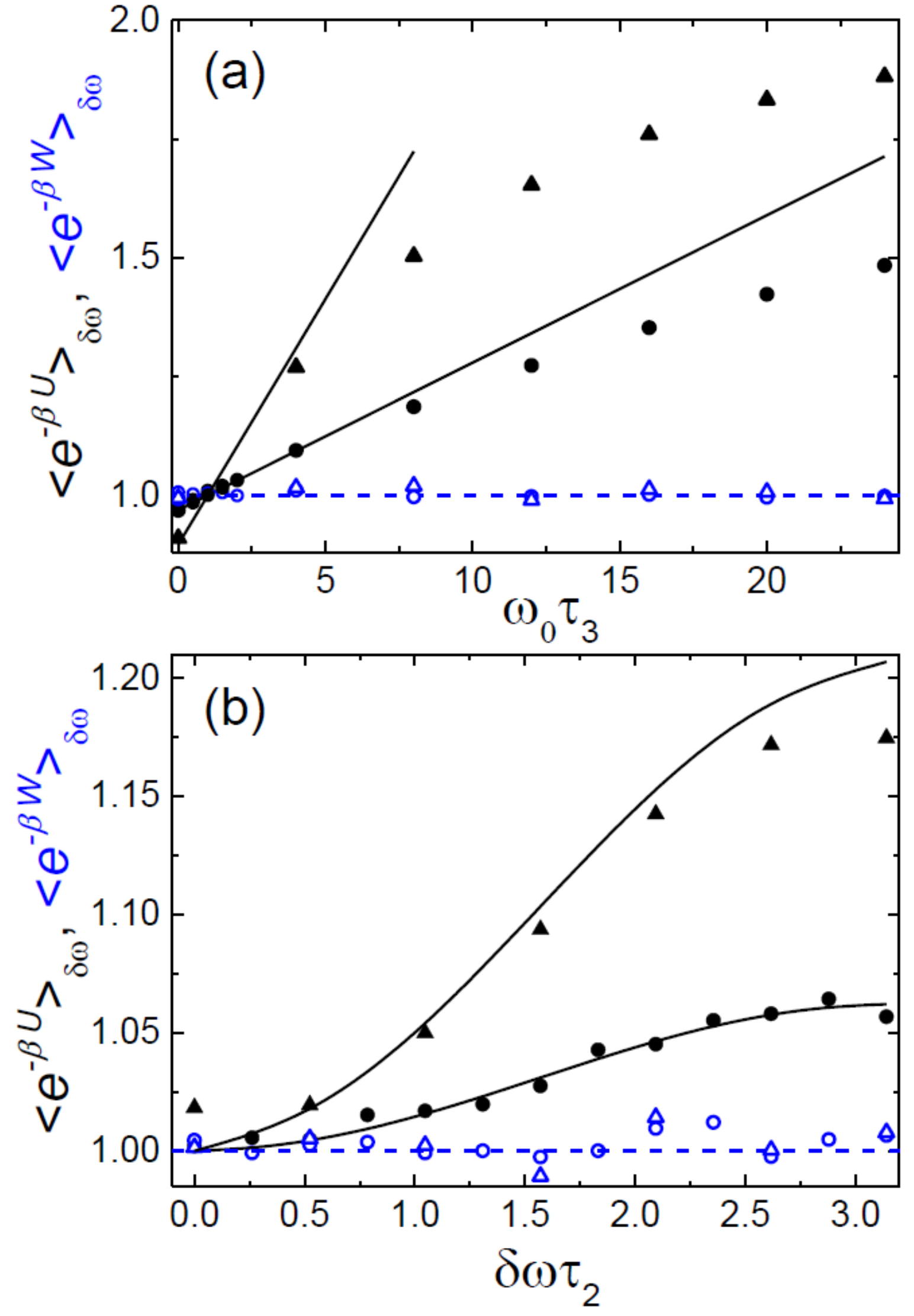}
\caption{Analytic expressions given in the text, compared to numerical results obtained by stochastic simulations, see Ref.~\cite{Hekking13} for details. The protocol is the one described in the text and in Fig.~\ref{fig1}. We assume $\beta\hbar\omega_0 =5$ and $\Gamma_\downarrow /\hbar \omega_0 = 0.03$ (circles), and  $\Gamma_\downarrow /\hbar \omega_0 = 0.1$ (triangles). In (a) $\omega_0\tau_1=\omega_0\tau_2=1$, $\omega_0\tau_3$ varies, and $\delta\omega=0$. The open symbols refer to $\langle e^{-\beta W}\rangle_{\delta\omega}$, and the filled ones to $\langle e^{-\beta U}\rangle_{\delta\omega}$. The solid lines represent the analytic predictions of Eq.~\eqref{ju1} for $\langle e^{-\beta U}\rangle_{\delta\omega}$ for the two values of $\Gamma_\downarrow$. In (b) $\langle e^{-\beta U}\rangle_{\delta\omega}$ is plotted for $\omega_0\tau_1=\omega_0\tau_2=\omega_0\tau_3 = 1$ against $\delta\varphi_2=\delta\omega\tau_2$ with other parameters as in (a) (solid lines and filled symbols). The results for $\langle e^{-\beta W}\rangle_{\delta\omega}$, indicated by the corresponding open symbols, are again concentrated around unity. In all cases we employed $10^7$ repetitions of the protocol for each data point.
\label{fig2}}
\end{figure}

Figure~\ref{fig2} shows the analytic predictions and numerical simulations of the quantities analyzed above for a chosen set of parameters as detailed in the figure caption. The offset $\delta \omega$ is assumed to be constant for each set of data. The numerical calculations are performed using a stochastic simulation with the QJ method, assuming ideal rotations. What one observes apart from the statistical scatter due to the finite number of repetitions in the averages, $10^{7}$ for each point, is that the analytic approximations follow closely the numerical results. Only for long $\tau_3$ intervals in (a), we see that the linear approximation for 
$\langle e^{-\beta U}\rangle_{\delta \omega}$ 
overestimates the deviation from unity. This is natural, as we have shown that for $\tau_3 \rightarrow \infty$, 
$\langle e^{-\beta U}\rangle_{\delta\omega} \approx 1.97 $ 
for the parameters in Fig.~\ref{fig2}, see Eq.~\eqref{hh2}. On the contrary, in all cases 
$\langle e^{-\beta W}\rangle_{\delta\omega}$ 
stays around unity within the statistical scatter. 

One obtains further insight of the differences between $U$ and $W$ by inspecting the actual probability distributions of internal energy and work under the given driving protocol. Both 
ratios $U/\hbar \omega_0$ and $W/\hbar \omega_0$ can obtain integer values.
Figure~\ref{fig3} shows the numerically calculated distributions for $U$ and $W$ under the same conditions as in Fig.~\ref{fig2} varying the delay time $\tau_3$ between the second rotation and the final measurement. There are several important conclusions to draw from the dependencies in Fig.~\ref{fig3}. Firstly, the probabilities 
$p_U(U)$
relax with time $\tau_3$ since $U$ measures the internal energy which changes after the non-equilibrium driving. On the contrary, the waiting period $\tau_3$ does not influence $p_W(W)$ 
since there is no work done after the second rotation pulse. These dependences are consistent with our basic expectations. Finally, we also recover the important Crooks fluctuation relation for work \cite{Crooks99} in the form 
\begin{equation} \label{crooks}
p_W(W)/p_W(-W) = e^{\beta W}
\end{equation} 
for those values of work that are within the reach in this situation (for $W/\hbar \omega_0=\pm 1$). On the contrary, the ratio $p_U(U)/p_U(-U)$ is not constant in $\tau_3$ and thus does not satisfy a fluctuation relation. 
\begin{figure}[t]
\centering
\includegraphics[width=.7\columnwidth]{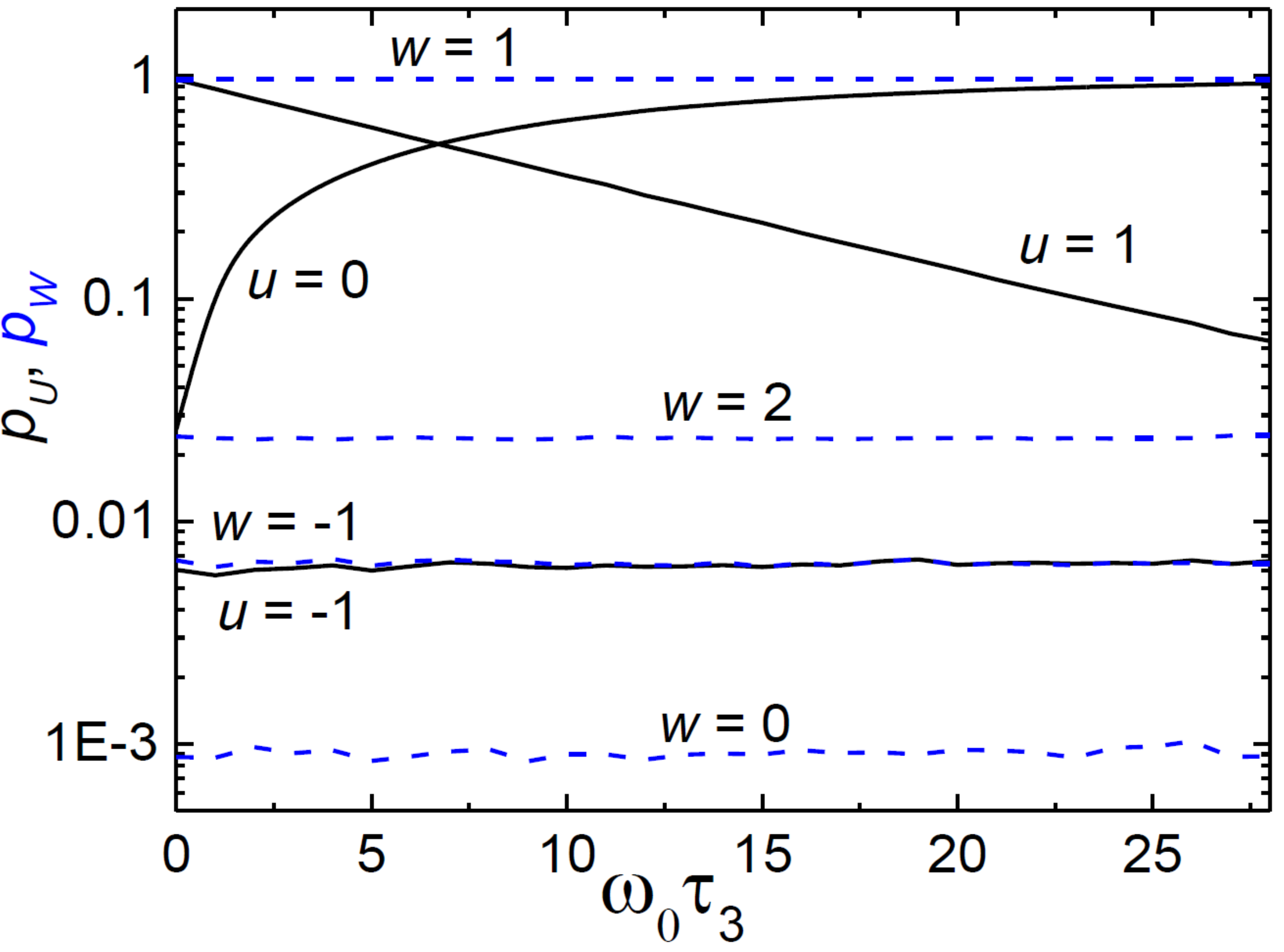}
\caption{Probabilities of different possible outcomes of $u \equiv U/\hbar\omega_0$ and $w \equiv W/\hbar\omega_0$. We assume the same protocol as before with $\beta\hbar\omega_0 =5$, $\Gamma_\downarrow /\hbar \omega_0=0.1$, $\delta\omega=0$, and $\omega_0\tau_1=\omega_0\tau_2=1$. We vary the delay time of the second measurement, $\tau_3$. The solid lines refer to $U$ and the dashed ones to $W$. We employed $3\cdot 10^5$ repetitions of the protocol for each data point.}
\label{fig3}
\end{figure}  

Neglecting time dependence of $\delta \omega$ during the time interval $\tau_2$ and assuming a Gaussian distribution of $\delta \omega$ with vanishing mean and width $\langle \delta\omega ^2 \rangle \equiv 2 \Gamma_\varphi^2$ we obtain
\begin{equation} \label{yu2}
\langle e^{-\beta U}\rangle -1=  [(\tau_3 - e^{-(\Gamma_\varphi \tau_2)^2} \tau_1]\,\Gamma_\Sigma \tanh^2(\beta\hbar\omega_0/2).
\end{equation} 
Therefore, $\Delta$ is directly related to $\Gamma_\varphi$, which is the decay rate for   $ \langle \cos(\delta\varphi_2) \rangle$.
This is the dephasing rate of the qubit during the time interval $\tau_2$.

A realistic source of decoherence caused by a low-frequency noise is discussed in the Supplemental Material, Sec.~2. We consider so-called random telegraph noise
created by a two-state dynamic degree of freedom in the environment, as well as $1/f$-noise produced by overlap of many degrees of freedom, see~\cite{Dutta81,*Kogan1996} for a general review and~\cite{NJP,*Paladino2014} for a review on qubits. Experiments~\cite{Nakamura02,*Astafiev04} on Josephson qubits have indicated that charged defects may be responsible for the $1/f$ noise. 

Another mechanism of decoherence is a back-action of the measuring device, see, e.g.,~\cite{Averin02,*Averin03}. In this case the decoherence rate can be related to the so-called \textit{accessible information}, i.e., the information gained in the measurement process, see, e.g.,~\cite{Pilgram2002,*Clerk2003} and references therein. If this decoherence mechanism is dominant, Eq.~\eqref{yu1} allows  relating thermodynamics of the qubit to the accessible information gained by the detector. 
Note, however, that the exact relationship between the decoherence and the accessible information depends on the detector type and the measurement protocol.

The decoherence and thermodynamics can be related to the mutual information also in the case when decoherence is due to interaction with fluctuators, but only when one is able to measure both the qubit and the fluctuators. The relationship between decoherence and  mutual information has been analyzed for a qubit interacting with a single fluctuator coupled to phonon environment~\cite{Wold2012}.

To summarize, we have studied the distribution of both the internal energy $U$ and the work $W$ done on a qubit in a driven non-equilibrium protocol under weak dissipation. Specifically we relate the non-equilibrium thermodynamic relations to decoherence rates. We demonstrate that the common fluctuation relations are satisfied for $W$, but not for $U$ which is the quantity measured in the two-measurement protocol, TMP. Our results allow one to evaluate the validity of, e.g., the Jarzynski equality from measurements of the  internal energy $U$ rather than of the total work $W$. From the opposite point of view, if the average $\langle e^{-\beta U}\rangle$ is measured, one can use the results for studying dephasing of the qubit via such a statistical measurement. 

The work has been supported partially by the Academy of Finland (projects 250280 and 272218), and the European
Union FP7 project INFERNOS (grant agreement 308850).


\appendix
\section{Dephasing and dissipation in qubit thermodynamics: Supplemental material}
\subsection{1. Jarzynski equality in the weak dissipation limit} \label{JE}

\begin{figure}[t]
\centering
\includegraphics[width=\columnwidth]{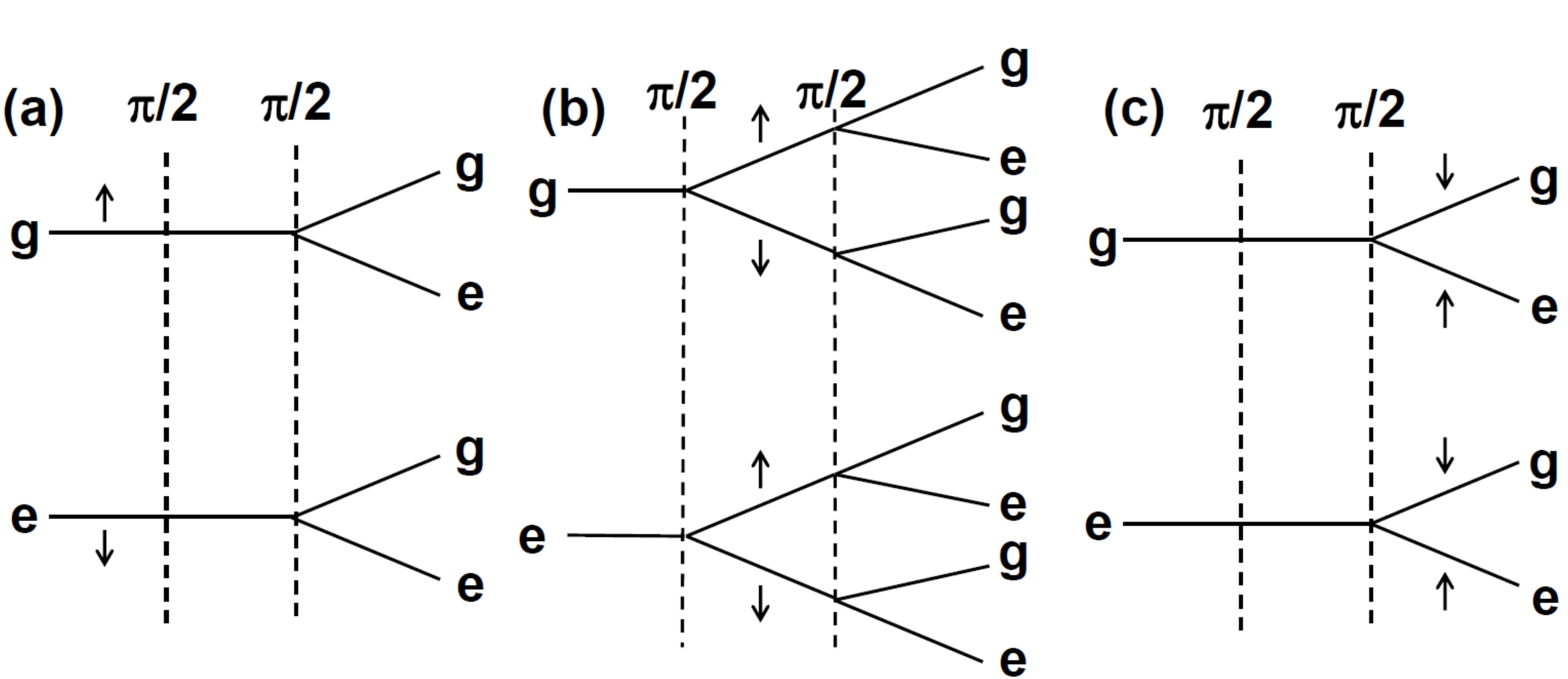}
\caption{Diagrams illustrating the one jump trajectories contributing to various $P_1\langle\cdot\rangle_1$ averages described in the text. On the left of each diagram, we indicate the result of the initial measurement ($g,e$), the vertical dashed lines indicate the rotations of the qubit and on the right we indicate the outcome of the final measurement ($g,e$). The relaxation and excitation events are indicated by down and up arrows, respectively. The jump occurs in the time interval $0 < t < \tau_1$ in (a), $\tau_1 < t < \tau_1+\tau_2$ in (b), and $\tau_1+ \tau_2 < t < \tau_1+\tau_2+\tau_3$ in (c). \label{fig_s}}
\end{figure}

Hereby we derive the expression for $\langle e^{-\beta U}\rangle$, Eq.~(11) of the main text, and prove that the  Jarzynski equality (JE)
\begin{equation} \label{je}
\langle e^{-\beta W}\rangle =1
\end{equation}
holds in the weak dissipation regime. 
In this limit we take into account trajectories which include at most one relaxation or excitation event, see Fig.~\ref{fig_s}, and obtain corrections to fluctuation relations linear in the transition rates. We may write the work exponent 
as
\begin{equation} \label{h2}
\langle e^{-\beta W}\rangle = P_0\langle e^{-\beta W}\rangle_0 + P_1 \langle e^{-\beta W}\rangle_1,
\end{equation}   
where $P_0,P_1$ are the probabilities of zero and one-photon processes, and $\langle e^{-\beta W}\rangle_0,\langle e^{-\beta W}\rangle_1$ are the averages for the corresponding processes. We will explicitly check the validity of JE up to this order. On the other hand, we have for the quantity 
$U$
measured in the two-measurement protocol (TMP),
\begin{equation} \label{h3}
\langle e^{-\beta U}\rangle = P_0\langle e^{-\beta U}\rangle_0 + P_1 \langle e^{-\beta U}\rangle_1
\end{equation}  
with the corresponding notations. Now for 
zero-photon
processes, $U=W$, thus $\langle e^{-\beta U}\rangle_0 = \langle e^{-\beta W}\rangle_0$.
Following the protocol outlined in Fig.~1 of the main text we find the amplitudes along the operation 
trajectories
of the qubit. If the system is in state $g$ just before the first $\pi/2$ rotation, it acquires amplitudes $$a_g(\tau_{1+})=1/\sqrt{2}, \ b_g(\tau_{1+})=1/\sqrt{2}$$ right after it. Similarly $$a_e(\tau_{1+})=-1/\sqrt{2}, \ b_e(\tau_{1+})=1/\sqrt{2}.$$ In the absence of jumps, the system evolves during the time interval $\tau_2$ such that according to Eq.~(5) of the main text, up to linear order in $\Delta \Gamma$,
\begin{eqnarray} \label{amp1}
a_g(t_{2-}) &=& - a_e(t_{2-}) = \frac{1}{\sqrt{2}}(1+\Delta \Gamma \tau_2/4)e^{i\delta\varphi_2/2}, \nonumber \\
b_g(t_{2-}) &=& \phantom{-}  b_e(t_{2-}) = \frac{1}{\sqrt{2}}(1-\Delta \Gamma \tau_2/4)e^{-i\delta\varphi_2/2},\nonumber
\end{eqnarray}
where $t_2 \equiv\tau_1+\tau_2$, $\delta \varphi_2 \equiv \int_0^{\tau_2} \delta \omega (t)\, dt$.
After the second $\pi/2$ rotation in this situation,
\begin{eqnarray} \label{amp2}
a_g(t_{2+}) &=& - b_e(t_{2+}) = i\sin(\delta \varphi_2/2)+\frac{1}{4}\Delta \Gamma \tau_2\cos(\delta\varphi_2/2) , \nonumber \\
b_g(t_{2+}) &=& -a_e(t_{2+}) = \cos(\delta \varphi_2/2)+\frac{i}{4}\Delta \Gamma \tau_2\sin(\delta\varphi_2/2) .\nonumber
\end{eqnarray}
Finally, the probabilities at the moment of the second measurement, $t_3\equiv\tau_1+\tau_2+\tau_3$, again assuming no jumps after the first rotation, are given by
\begin{eqnarray} \label{amp3}
&& |a_g(t_3)|^2 = \sin^2(\delta \varphi_2/2)[1+\Delta\Gamma \tau_3\cos^2(\delta\varphi_2/2)],\nonumber \\&&
|a_e(t_3)|^2 =\cos^2(\delta \varphi_2/2)[1+\Delta\Gamma \tau_3 \sin^2(\delta\varphi_2/2)],\nonumber
\end{eqnarray}
with $|b_{g,e}(t_3)|^2 =1-|a_{g,e}(t_3)|^2$.

With the help of the evolution of the amplitudes, we can construct the averages. In particular for no-jump trajectories we may write
\begin{eqnarray} \label{nojump}
&&  P_0\langle e^{-\beta U}\rangle_0 =P_0\langle e^{-\beta W}\rangle_0 =  \nonumber \\
&& \quad  p_g e^{-d_g(t_3)} (|a_g(t_3)|^2e^0 +|b_g(t_3)|^2 e^{-\beta\hbar\omega_0})
\nonumber\\
&& \quad \quad
+p_e e^{-d_e(t_3)}(|a_e(t_3)|^2e^{\beta\hbar\omega_0} +|b_e(t_3)|^2 e^0). 
\end{eqnarray}
Here $p_g=1-p_e=(1+e^{-\beta\hbar\omega_0})^{-1}$,
$$d_i (\tau) \equiv \int_0^{\tau} dt \, (\Gamma_\uparrow |a_i(t)|^2 + \Gamma_\downarrow |b_i(t)|^2).$$
Taking only contributions yielding up to linear corrections in $\Gamma$'s, we obtain
\begin{eqnarray} \label{nojump2}
&& P_0\langle e^{-\beta U}\rangle_0 =P_0\langle e^{-\beta W}\rangle_0 = 1 - 
\nonumber \\
&& p_g\Gamma_\downarrow\tau_1[\cos^2(\delta\varphi_2/2)(1+e^{-2\beta\hbar\omega_0})+2\sin^2(\delta\varphi_2/2)e^{-\beta\hbar\omega_0}]\nonumber\\&& \quad \quad 
-\frac{1}{2}\Gamma_\Sigma\tau_2-2p_g\Gamma_\downarrow\tau_3 e^{-\beta\hbar\omega_0}.
\end{eqnarray}
Here $\Gamma_\Sigma \equiv \Gamma_\downarrow + \Gamma_\uparrow$.
For the single-jump trajectories, we follow the diagrams in Fig.~\ref{fig_s} in this supplemental material to obtain again the contributions up to linear order in $\Gamma$'s. As an example, for the diagrams in Fig.~\ref{fig_s}~(a) we obtain
\begin{eqnarray} \label{firstaU}
&&P_{1,a} \langle e^{-\beta U}\rangle_{1,a} = p_g\Gamma_\uparrow\tau_1(|a_e(t_3)|^2 e^0 + |b_e(t_3)|^2 e^{-\beta\hbar\omega_0})\nonumber\\
&& + p_e\Gamma_\downarrow\tau_1(|a_g(t_3)|^2 e^{\beta\hbar\omega_0} + |b_g(t_3)|^2 e^0)\nonumber \\
&&=p_g\Gamma_\downarrow\tau_1[\sin^2(\delta\varphi_2/2)\left(1+e^{-2\beta\hbar\omega_0}\right)  \\ && \quad \quad 
 +2\cos^2(\delta\varphi_2/2)e^{-\beta\hbar\omega_0}] ,\nonumber
\end{eqnarray}
and
\begin{eqnarray} \label{firstaW}
&&P_{1,a} \langle e^{-\beta W}\rangle_{1,a} = p_g\Gamma_\uparrow\tau_1(|a_e(t_3)|^2 e^{\beta\hbar\omega_0} + |b_e(t_3)|^2 e^0)
\nonumber\\&& + p_e\Gamma_\downarrow\tau_1(|a_g(t_3)|^2 e^0 + |b_g(t_3)|^2 e^{-\beta\hbar\omega_0})
\nonumber \\&&=p_g\Gamma_\downarrow\tau_1[\cos^2(\delta\varphi_2/2)(1+e^{-2\beta\hbar\omega_0})
 \\ && \quad \quad 
+2\sin^2(\delta\varphi_2/2)e^{-\beta\hbar\omega_0}].\nonumber
\end{eqnarray}
In these equations $\Gamma_{\downarrow,\uparrow} \tau_1 \approx 1-e^{-\Gamma_{\downarrow,\uparrow} \tau_1}$ is the probability of a jump within the time interval $\tau_1$.

For the diagrams in Fig.~\ref{fig_s}~(b) we obtain
\begin{eqnarray} \label{firstbU}
&& P_{1,b} \langle e^{-\beta U}\rangle_{1,b} = \nonumber\\&&
p_g\Gamma_\uparrow |a_{g,\pi /2}|^2 \tau_2(|a_{e, \pi /2}|^2 e^0 + |b_{e,\pi /2}|^2 e^{-\beta\hbar\omega_0})\nonumber\\&&
+ p_g\Gamma_\downarrow |b_{g,\pi /2}|^2 \tau_2(|a_{g, \pi /2}|^2 e^0 + |b_{g,\pi /2}|^2 e^{-\beta\hbar\omega_0}) \nonumber\\&&
+ p_e\Gamma_\uparrow |a_{e,\pi /2}|^2 \tau_2(|a_{e, \pi /2}|^2 e^{\beta\hbar\omega_0} + |b_{e,\pi /2}|^2 e^0) \nonumber\\&&
+ p_e\Gamma_\downarrow |b_{e,\pi /2}|^2 \tau_2(|a_{g, \pi /2}|^2 e^{\beta\hbar\omega_0} + |b_{g,\pi /2}|^2 e^0).
\end{eqnarray}
Here $|a_{g,\pi /2}|^2 =|a_{e,\pi /2}|^2 = |b_{g,\pi /2}|^2 = |b_{e,\pi /2}|^2 =1/2$ are the probabilities after a $\pi/2$ pulse when the system starts from an eigenstate. Inserting these values we obtain
\begin{equation} \label{U1b}
P_{1,b} \langle e^{-\beta U}\rangle_{1,b} =\frac{1}{2}\Gamma_\Sigma \tau_2.
\end{equation}
Similarly,
\begin{eqnarray} \label{firstbW}
&& P_{1,b} \langle e^{-\beta W}\rangle_{1,b} = \nonumber\\&&
p_g\Gamma_\uparrow |a_{g,\pi /2}|^2 \tau_2(|a_{e, \pi /2}|^2 e^{\beta\hbar\omega_0} + |b_{e,\pi /2}|^2 e^0)\nonumber\\&&
+ p_g\Gamma_\downarrow |b_{g,\pi /2}|^2 \tau_2(|a_{g, \pi /2}|^2 e^{-\beta\hbar\omega_0} + |b_{g,\pi /2}|^2 e^{-2\beta\hbar\omega_0}) \nonumber\\&&
+ p_e\Gamma_\uparrow |a_{e,\pi /2}|^2 \tau_2(|a_{e, \pi /2}|^2 e^{2\beta\hbar\omega_0} + |b_{e,\pi /2}|^2 e^{\beta\hbar\omega_0}) \nonumber\\&&
+ p_e\Gamma_\downarrow |b_{e,\pi /2}|^2 \tau_2(|a_{g, \pi /2}|^2 e^0 + |b_{g,\pi /2}|^2 e^{-\beta\hbar\omega_0}),
\end{eqnarray}
yielding
\begin{equation} \label{W1b}
P_{1,b} \langle e^{-\beta W}\rangle_{1,b} =\frac{1}{2}\Gamma_\Sigma \tau_2.
\end{equation}

For the diagrams in Fig.~\ref{fig_s}~(c), we have
\begin{eqnarray} \label{firstcU}
&& P_{1,c} \langle e^{-\beta U}\rangle_{1,c} = \nonumber\\&&
p_g(\Gamma_\uparrow |a_g(t_{2+})|^2 \tau_3 e^{-\beta\hbar\omega_0} + \Gamma_\downarrow |b_g(t_{2+})|^2 \tau_3 e^0)\nonumber\\&&
+p_e(\Gamma_\uparrow |a_e(t_{2+})|^2 \tau_3 e^0 + \Gamma_\downarrow |b_e(t_{2+})|^2 \tau_3 e^{\beta\hbar\omega_0}),
\end{eqnarray}
yielding
\begin{equation} \label{U1c}
P_{1,c} \langle e^{-\beta U}\rangle_{1,c} = p_g\Gamma_\downarrow\tau_3(1+e^{-2\beta\hbar\omega_0}).
\end{equation}
Similarly we obtain for $W$ for the diagrams in Fig.~\ref{fig_s}~(c)
\begin{eqnarray} \label{firstcW}
&& P_{1,c} \langle e^{-\beta W}\rangle_{1,c} = \nonumber\\&&
p_g(\Gamma_\uparrow |a_g(t_{2+})|^2 \tau_3 e^0 + \Gamma_\downarrow |b_g(t_{2+})|^2 \tau_3 e^{-\beta\hbar\omega_0})\nonumber\\&&
+p_e(\Gamma_\uparrow |a_e(t_{2+})|^2 \tau_3 e^{\beta\hbar\omega_0} + \Gamma_\downarrow |b_e(t_{2+})|^2 \tau_3 e^0),
\end{eqnarray}
which gives
\begin{equation} \label{W1c}
P_{1,c} \langle e^{-\beta W}\rangle_{1,c} = 2p_g\Gamma_\downarrow\tau_3 e^{-\beta\hbar\omega_0}.
\end{equation}

Combining Eqs. \eqref{nojump2}, \eqref{firstaW}, \eqref{W1b} and \eqref{W1c}, we obtain up to the first order in $\Gamma$'s, 
\begin{equation} \label{yu0}
\langle e^{-\beta W}\rangle_{\delta\omega} \equiv P_0\langle e^{-\beta W}\rangle_0 +\sum_{i=a,b,c}P_{1,i}\langle e^{-\beta W}\rangle_{1,i} =  1 ,
\end{equation} 
where the subscript $\delta\omega$ emphasizes that we average over all realizations with a fixed level spacing $\omega_0+\delta\omega$. This result implies that the Jarzynski equality, $\langle e^{-\beta W}\rangle =1$, is valid for any distribution of $\delta \omega$. Similarly, combining Eqs.~\eqref{nojump2}, \eqref{firstaU}, \eqref{U1b} and \eqref{U1c}, we obtain Eq.~(11) of the main text.
\subsection{2. Decoherence due to low-frequency noise}

Here we will discuss
a realistic source of decoherence caused by a low-frequency noise. A part of these fluctuations typically has a $1/f$ spectrum and is referred to as $1/ f $ noise. 
Such noise is generic for all disordered materials (for a review see, e.\,g.,~\cite{Dutta81} and references therein). It is also common in single-electron and other tunneling devices, see, e.\,g.,~\cite{Zorin96}. Experiments~\cite{Nakamura02,*Astafiev04} on Josephson qubits have indicated that charged impurities may be responsible for the $1/f$ noise. 

One of the most common sources of low-frequency noise is the rearrangement of dynamic two-state defects, {\em fluctuators}, see, e.~g.,~\cite{Kogan1996} and references therein. Random switching of a fluctuator  between its two metastable states (1 and 2)  produces random telegraph noise. The process is characterized by the switching rates $\gamma_{12}$ and $\gamma_{21}$ for the transitions $1\to 2$ and $2\to 1$. Only the fluctuators with energy splitting $E \lesssim T$, contribute to the dephasing since the fluctuators with large level splitting are frozen in their ground states (we measure temperature in the energy units).  As long as $E<T$ the rates $\gamma_{12}$ and $\gamma_{21}$
are close in magnitude, and without loss of generality one can assume that $\gamma_{12} \approx \gamma_{21} \equiv \gamma$.\, i.\,e., the fluctuations can be described as a \textit{random telegraph process}, for reviews see~\cite{Kogan1996,RTP}. A set of random telegraph fluctuators with exponentially broad distribution of relaxation rates, $\gamma$, produces noise with $1/f$ power spectrum at $\gamma_{\min} \ll \omega=2\pi f \ll \gamma_0$. Here $\gamma_{\min}$ is the switching rate of the ``slowest'' fluctuator while $\gamma_0$ is the maximal switching rate for $E=T$. 
Random telegraph noise has been observed in numerous nanodevices based both on semiconductors, normal metals, and superconductors~\cite{RTN}.

For evaluating role of fluctuators in dephasing, we will use a simple classical model within which one can analyze exactly the qubit response to typical manipulation protocols.  According to  this  model, the quantum system -- qubit -- interacts with a set of two-level entities, see~\cite{NJP,Paladino2014} for a review. The latter fluctuate stochastically between their states due to interaction with a thermal bath, which may be not directly coupled to the qubit. 

We start from the essentially non-Gaussian situation when there is a single fluctuator coupled to the qubit. Following~\cite{NJP}, we assume that the fluctuator is described by the Hamiltonian $\mathcal{H}_F = (E/2) \tau_z$ where $\tau_z$ is the Pauli matrix of the fluctuator and  $ E=\sqrt{\Delta^2+\Lambda^2}$ is its energy splitting. The latter depends on the diagonal splitting $\Delta$ between the fluctuator's states and their tunneling coupling $\Lambda$.

This fluctuator switches between its states due to the interaction with environmental bosons (phonons or electrical fluctuations), the switching rate being~\cite{NJP}
\begin{equation} \label{5}
\gamma =(\Lambda/E)^2 \gamma_0 (E).
\end{equation}
A typical estimate for $\gamma_0$ is  $T^3/\hbar T_*^2$, where $T_*\sim$ 20 K.

Following the approach outlined in~\cite{NJP}, we represent $\delta \omega (t)$ as a random telegraph process, $\chi(t)$, as  $\delta \omega (t)=v\chi(t)$. Here $\chi (t)$ switches between the values $\pm 1$ at random times, distributed according to the Poisson distribution, $\langle \chi(t)\chi(0) \rangle = e^{-2\gamma t}$.
The coupling constant $v$, in general, depends on the operation point of the qubit,
$v\propto (\Delta_q/\hbar \omega_0) (\Delta/E)$, so the mechanism can be verified by changing this point. 
Here $\Delta_q$ is the tunneling splitting of the qubit.
\begin{figure}[b]
\centering
\includegraphics[width=.4\columnwidth]{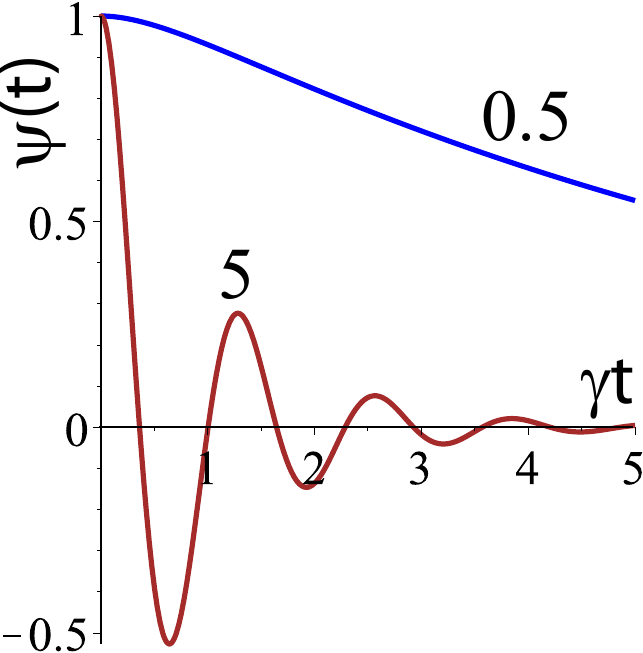}
\caption{Plots of the function $\psi(v,\gamma|t)$
given by
Eq.~\eqref{psi1} for $v/\gamma =0.5$ (blue curve) and  $v/\gamma =5$ (brick-red curve) . \label{fig3}}
\end{figure}

The average  $\psi (v,\gamma|\tau_2) \equiv \langle \cos(\delta \varphi_2)\rangle$ reads as~\cite{NJP}
\begin{equation} \label{psi1}
  \psi(v,\gamma|\tau_2) = \frac{e^{-\gamma \tau_2}}{2\mu}\left[(\mu +1)e^{\gamma \mu \tau_2}+ (\mu -1)e^{-\gamma \mu \tau_2}\right] \! \!
\end{equation}
where $\mu \equiv  \sqrt{1-(v/\gamma)^2}$.
At $v \ll \gamma$, 
\begin{equation} \label{8}
\psi(v,\gamma|\tau_2) \sim e^{-\Gamma_\varphi \tau_2}\, , \quad \Gamma_\varphi =v^2/2\gamma\, .
\end{equation}
This behavior is typical for the so-called \textit{motional narrowing} of the spectral lines in magnetic resonance~\cite{KlauderAnderson}. 
At $v \gg \gamma$,
\begin{equation} \label{9}
\psi(v,\gamma|\tau_2) \sim e^{-\gamma \tau_2}\cos(v \tau_2)\, ,
\end{equation}
i.e., we have beating between the qubit's  levels $E_0 \pm v $. These dependences are shown in Fig.~\ref{fig3}.

Now we consider the case of many fluctuators producing $1/f$ noise. We will use the same approach as in~\cite{NJP} and assume that the random processes of different fluctuators are not correlated and their total number $N \gg 1$. 
Then $\langle \cos(\delta \varphi_2) \rangle$
can be expressed as $e^{-\mathcal{K} (\tau_2)} $ where 
\begin{equation} \label{kt1}
\mathcal{K}(\tau_2)=\int dv\, d \gamma\, \mathcal{P}(v,\gamma)\left[1-\psi(v,\gamma|\tau_2)\right]\, .
\end{equation}
Here $\psi( v,\gamma|t)$ is given by Eq.~\eqref{psi1}, while $\mathcal{P}(v,\gamma)$ is the distribution of the coupling parameters, $v$, and the fluctuator's switching rates, $\gamma$. 

To specify the distribution function one has to assume a model for fluctuators.  An overview of different models is given in Refs.~\cite{NJP,Paladino2014}.
To formulate the results let us introduce a typical coupling strength $\eta$ as the interaction strength at the average distance between fluctuators with the interlevel spacing of $T$. Here we will briefly discuss the case when fluctuators are uniformly distributed in the space around qubit and the interaction between the qubit and a fluctuator decays $\propto 1/r^3$. In this case~\cite{laikhtman,ags}
\begin{equation}\label{18}
\mathcal{K}(\tau_2)\approx \eta\cdot\left\{ \begin{array}{lll}
\gamma_0 \tau_2^2& \text{for} &\gamma_0 \tau_2 \ll 1 \, ; \\
\tau_2\ln \gamma_0 \tau_2& \text{for} &\gamma_0 \tau_2 \gg 1\, .
\end{array}\right.
\end{equation}
The results for the first limiting case coincide with the Gaussian decay, Eq.~(16) of the main text, with $\Gamma_\varphi \approx \sqrt{\eta \gamma_0}$. This result has a clear physical
meaning: the decoherence occurs only provided that at least one of the fluctuators flips. Each flip provides a contribution $\sim \eta \tau_2$ to the phase, while $\gamma_0 \tau_2  \ll 1$ is a probability for a flip during the observation time.  The result for $\gamma_0 \tau_2\gg 1$ is less intuitive since in this region the dephasing is non-Markovian, see~\cite{laikhtman} for more details.

It is important that at large observation times, $\tau_2  \gg \gamma_0^{-1}$, the decoherence is dominated by few  \emph{optimal} fluctuators. The distance $r_{\scriptsize \textrm{opt}}(T)$, between
the optimal fluctuators and the qubit is determined by the condition
  \begin{equation}
    \label{eq:aux1}
    v(r_{\text{opt}}) \approx \gamma_0(T)\, .
  \end{equation}
This estimate emerges naturally from the behavior of the decoherence in the limiting cases $v \gg \gamma$ and $v \ll \gamma$. For strong coupling the fluctuators are slow and the qubit's behavior is determined by quantum beatings between the states with $E\pm v$. Accordingly, the decoherence rate is of he order of $\gamma$. In the opposite case, as we already discussed, the decoherence rate is $\propto v^2/2\gamma$. Matching these two limiting cases one arrives at the estimate~(\ref{eq:aux1}).

Let us discuss possible implications of the results for the suggested TMP. We start with the case of a single strong fluctuator.
In principle, one can tune the qubit (i.e., the manipulation frequency)  to $\langle \delta \omega  \rangle  =0$ using the TMP.   Note, however, that such a tuning would be temperature-dependent since $\langle \delta \omega \rangle$ generally depends on temperature through the fluctuator's occupation numbers. Therefore, it may happen that tuning should be made for each temperature.

As we have discussed, the behavior of $\psi(v,\gamma|\tau_2)$ depends on one dimensionless parameter, $v/\gamma_0(T)$. In both cases theory predicts \text{exponential} decay, that corresponds to the Lorentzian spectrum. However, the decay rates in the different regimes are very different depending on temperature. In the case of a ``weak" fluctuator, the decay rate is $\Gamma_\varphi =v^2/2\gamma$, cf. with Eq.~\eqref{8}, where $\gamma \propto \gamma_0(T)$, see Eq.~\eqref{5}, is increasing with temperature (the typical dependence is $\propto T^3$). 

At low temperatures one can expect the parameter $v/\gamma_0(T)$ to become large. In this regime one can expect damped  beatings with frequency $v$, see Eq.~\eqref{9}, the decay rate being $\gamma$. In principle, one can find the coupling strength, $v$, from the beatings' frequency. If a crossover between the regimes is experimentally feasible, then one can be sure that at the crossover point the fluctuator is \textit{optimal} in the sense of Eq.~\eqref{eq:aux1}, i.e.,~it is the most harmful.

If the decoherence is produced by many fluctuators ($1/f$-noise) we have to study Eq.~\eqref{18}, which also predicts a temperature-dependent crossover at $\gamma_0(T) \tau_2 \approx 1$. At low temperatures and relatively small delay times, $\gamma_0(T)\tau_2 \ll 1$, the predicted decay is Gaussian. The predicted decay rate in this case, $\Gamma_\phi \approx \sqrt{\eta (T) \gamma_0 (T)} \propto T^2$.
Note that $\eta$ is proportional the number of fluctuators with $E\lesssim T$, therefore $\eta(T) \propto T$. 
At higher temperatures a crossover to the case $\gamma_0(T)\tau_2 \gg 1 $ takes place. In this - non-Markovian - regime, the predicted decay is exponential, the decay rate being $\eta (T) \propto T$.

The above discussion holds for the situation when the fluctuators are evenly distributed in a 3D device and the interaction between the qubit and the fluctuators decays $\propto 1/r^3$.  If the fluctuators are formed close to some low-dimensional surface, 
 then the situation can be more diverse. 
Typically, in that case a small group of fluctuators becomes dominant, the decoherence is non-Gaussian, and pronounced mesoscopic fluctuations of the decoherence time can be expected. Generally, the behavior of the decay can be accounted assuming that there is one or few strong fluctuators and applying expressions for a single fluctuator, see more details in~\cite{NJP,Paladino2014}.

\end{document}